\documentclass[preprint,epsfig,graphicx,showpacs,eqsecnum,aps,amsmath,mathrsfs,eufrak]{revtex4}
\usepackage{graphicx}
\usepackage{epsfig}

\newcommand{\be}{\begin{equation}}
\newcommand{\ee}{\end{equation}}
\newcommand{\bea}{\begin{eqnarray}}
\newcommand{\eea}{\end{eqnarray}}
\begin{document}
\draft
\title{Collective dipole excitations in sodium clusters}
\author{A. A. Raduta$^{a,b)}$, R. Budaca $^{b)}$,
and Al. H. Raduta$^{b)}$ }

\address{$^{a)}$Department of Theoretical Physics,
Bucharest University, POB MG11, Romania}
\address{$^{b)}$Institute of Physics and Nuclear Engineering, Bucharest,
POB MG6, Romania}
\date{\today}
\begin{abstract}
Some properties of small and medium sodium clusters are described
within the RPA approach using a projected spherical single particle basis. The oscillator strengths calculated with a Schiff-like dipole transition operator and folded with Lorentzian functions are used to calculate the photoabsorbtion cross section spectra. The results are further employed to establish the dependence of the plasmon frequency on the number of cluster components. Static electric polarizabilities of the clusters excited in a RPA dipole state are also calculated.
 Comparison of our results with the corresponding experimental data show an overall good agreement. 
\end{abstract}
\narrowtext
\maketitle

\renewcommand{\theequation}{1.\arabic{equation}}
\section{Introduction}
\label{sec:level1}
 Although the field of metallic clusters is very old, it became very attractive for both theoreticians and experimentalists only since 1984,
when the pioneering paper of Knight et al \cite{Knight} pointed out the
electronic shells in alkali-metallic clusters.  
Some notable contributions in the new era of the field have been reviewed  
by several authors \cite{Gal,Heer,Brack1,Kresin}.

The clusters which are most interesting to be studied seem to be those with a
moderate number of atoms. Indeed, for such systems neither statistical models
\cite{Cini}
nor ${\it ab-initio}$ quantum-chemical methods \cite{Bonac} are justified.
Instead, the mean field approach is vastly used. Several solutions 
defining the mean field for the single particle motion have been employed
along the years. Among them, three procedures are to be distinguished:
i) solving the Kohn-Sham equations \cite{Kohn}; 
ii) assuming that the positive charge
of the ionic core is uniformly distributed in a sphere of radius R. This 
is known in the literature as jellium hypothesis \cite{Mart,Beck}; iii) postulating the average
potential \cite{Nish}.  

The researchers in this field payed attention especially to spherical clusters since
to this shape are  associated notions like shell structure or magic numbers. 
However, there are some features like the detailed structure of the abundance
spectrum \cite{Chou}, or the split of plasmon energies \cite{Voll,Selby1,Brec,Fall,Yan,Yan1,Kos},
which cannot be explained assuming a
spherical symmetry for the mean field. Moreover, measuring the split of the plasmon energy one can
 get information about the cluster deformation. 

The first paper devoted to the
deformed clusters was due to Clemenger and published in 1985 \cite{Clem}. 
The author adapts the Nilsson model formulated for nuclear systems \cite{Nils}, 
by ignoring the spin-orbit term. The resulting model is, therefore, referred to as 
Clemenger-Nilsson (CN) model.
The CN model was very successful in explaining several properties which
depend on the cluster shape and which could not be described within
a formalism using a mean field with spherical symmetry.
 The model is suitable to describe the
single particle properties in the intrinsic frame and especially for
strong coupling regime when the wave function of the whole system can be
factorized into an intrinsic part and a Wigner function accounting for the
rotational degrees of freedom. For clusters exhibiting an axial symmetry the projection of angular momentum on the symmetry axis, denoted by $K$ is a good quantum number. Of course the angular momentum itself is not a good quantum number. Having in view the fact that the measurements are achieved in the laboratory frame where the rotation symmetry is valid  we have to pay attention to this feature.
Indeed, there are many properties which are very sensitive to the change
of angular momentum of the system.
Moreover in most cases "K" is not a good quantum number and therefore the
factorization mentioned above is not possible. The typical case of this
kind is that of systems with triaxial shape. A many body treatment of such
situations would require a subsequent projection of the angular momentum.
 Such an operation is technically very difficult to be achieved and to
our knowledge up to now only approximate solutions have been adopted.
In a previous publication, one of us (A. A. R.) proposed a solution for
constructing a single particle basis with good spherical symmetry and
depending on deformation \cite{Rad}. In the quoted work
 the model ability to
account in a realistic fashion for the main features of the deformed
clusters has been successfully tested. Thus,  the cluster shape, the magic numbers,
and supershell effects have been determined and a good agreement with the
data as well as with the previous theoretical results have been obtained. For example our results concerning the supershell structure are consistent with the picture described in the work of Nishioka et al. \cite{Nish} as well as with the experimental data \cite{Pedersen,Martin}.

In a subsequent paper we  continued the exploration of the  deformed cluster
properties, within the projected spherical basis introduced in Ref.
\cite{Rad}. The single particle symmetry is actually the mean field symmetry. This feature is an important property related with the charge distribution of the valence electrons. Such a structure becomes very important when 
one studies the response of the cluster to the action of an external electromagnetic
field.

Guided by a possible parallelism between  atomic clusters and nuclear systems pointed out in 
Ref.\cite{Kosk1,Rigo}, in Ref.\cite{Raduta} one of us (A.A.R.) studied, in collaboration, properties like
skin structure, empty center, hard center, cluster
subsystems and halo behavior. 
Such properties have been  seen, indeed,  in the structure of the charge density function.
Based on phenomenological arguments, quantitative results for polarizability and plasmon frequency have been derived.

In this paper we study the collective many body properties of light and medium sodium clusters. The paper is structured as follows. In Section II we briefly review the projected single particle basis. In Section III the basic equations specific to the particle-hole ($ph$) Random Phase Approximation (RPA) are written down.
The RPA wave functions are used in Section IV, to treat the electric dipole transitions and the photoabsorbtion cross section.
In the framework of phRPA the expressions for system polarizability are  analytically derived, in Section V.
Numerical applications for Na clusters with the number of components varying from six to forty have been performed and the results are described in Section VI.
The final conclusions are summarized in Section VII.

\renewcommand{\theequation}{2.\arabic{equation}}
\section{Projected spherical single particle basis}
\label{sec: level2}
We restrict our considerations to the energy domain of laser beam
experiments (i.e. optical domain) where only the valence electrons may be
excited and de-localized, the remaining ones defining the atomic core.
Under these circumstances we could study those atomic cluster properties 
which are mainly determined by the valence electrons. The picture is even
more simplified if the cluster building block is an alkali-metal. In this
case the interacting system of electrons and positively charged ions is 
replaced by a system of interacting electrons moving in a mean field which
accounts for the influence of the ionic core on the single particle
motion. Each atom of a given cluster is represented by one valence electron. 

In a  previous publication \cite{Rad}, the mean field for the valence electrons
was defined with the help of a model Hamiltonian associated to the
particle-core interacting system.

\bea
H&=&\frac{p^2}{2m}+\frac{m\omega^2_0r^2}{2}-D\left(l^2-\langle l^2 \rangle\right)
+H_c-
m\omega^2_0r^2\sum\limits_{\lambda=0,2}\;\sum_{-\lambda\leq\mu\leq\lambda}
\alpha^*_{\lambda \mu}Y_{\lambda\mu}
\nonumber\\
&\equiv& H_p+H_c+H_{pc},
\eea
where $\alpha_{\lambda\mu}$ are $\lambda$-pole shape variables defining the deformed
ionic  core
through the surface equation:

\begin{equation}
R=R_0\left(1+\sum\limits_{\lambda=0,2}\sum_{-\lambda\leq\mu\leq \lambda}
\alpha^*_{\lambda\mu}Y_{\lambda\mu}\left(\theta,\phi\right)\right).
\end{equation}
The volume conservation condition allows us to relate the monopole
and quadrupole coordinates:

\be
\alpha_{00}=-\frac1{\sqrt{4\pi}}\sum\limits_{-2\leq\mu\leq2}|\alpha_{2\mu}|^2.
\ee
For what follows it is convenient to introduce the boson operators $b^{\dagger}_{2\mu}$:
\label{cantr}
\be
\alpha_{2\mu}=\frac1{k\sqrt{2}}\left(b^{\dagger}_{2\mu}+(-)^{\mu}b_{2-\mu}\right);~~~
\pi_{2\mu}=\frac{ik}{\sqrt{2}}\left(-b_{2\mu}+(-)^{\mu}b^{\dagger}_{2-\mu}\right).
\ee
The core subsystem is considered to be a harmonic quadrupole boson Hamiltonian:
\be
H_{c}=\omega_c\sum_{\mu}b^{\dagger}_{2\mu}b_{2\mu}.
\ee

Let us consider the coherent state
\be
|\Psi_c\rangle =exp\left[d\left(b^{\dagger}_{20}-b_{20}\right)\right]|0\rangle ,
\ee
where $|0\rangle$  denotes the vacuum state of quadrupole bosons.
This wave function is a coherent state:
\be
b_{2m}|\Psi_{c}\rangle=\delta_{m0}d|\Psi_{c}\rangle
\ee
and describes the ground state of a deformed quadrupole boson Hamiltonian. Moreover, the expected value of the static quadrupole moment in the state $|\Psi_c\rangle$, is proportional to $d$. Due to this feature, $d$ is referred to as {\it the deformation parameter}.

Note that averaging $H$ with  $|\Psi_c\rangle$, we obtain a deformed single particle mean field
which is similar to that used by Clemenger in Ref. \cite{Clem}. On the other hand,  averaging  $H$
with an eigenstate $|nlm\rangle$ of $H_p$, one arrives at a deformed quadrupole boson Hamiltonian which admits $|\Psi_c\rangle$  as ground state if a suitable deformation parameter is chosen.
These properties suggest that the particle-core system wave function might be described by a transformation applied to the product function with the factors $|nlm\rangle$ and $|\Psi_c\rangle$. 

The deformation of the single particle mean field in the Clemenger's model, $\delta$, might be related to the deformation parameter $d$ involved in the coherent state defined above.
Aiming at this goal, we require that the average of $H_{pc}$ with $|\Psi_{c}\rangle$ is identical to the deformed single particle potential from the Clemenger's Hamiltonian. This supplies us with the relation
\be
\frac{d}{k}=\sqrt{\frac{2\pi}{45}}\left(\Omega_{\perp}^{2}-\Omega_{z}^{2}\right),
\ee
where $\Omega_z$ and $\Omega_{\perp}$ denote the frequencies along and perpendicular to the symmetry axis, respectively. This relation
 yields a simple equation for the two deformation parameters:
\be
0.693k\delta=d.
\ee
In our calculations, the adopted value for  the constant $k$, defining the canonical transfromation (2.4), is 9.77. 
The mean field defines an intrinsic frame of reference for electrons. In the laboratory frame the system is described by states having good angular momentum, due to the rotation symmetry of the model Hamiltonian. A basis in the laboratory frame for the composite system, of electrons and core, may be obtained by diagonalizing $H$ in a particle-core product basis, with components of definite angular momenta. However, to use such a basis in a Random Phase Approximation (RPA) formalism is quite a tedious task.

A great simplification is obtained if, instead, we use a projected spherical single particle basis.
Obviously, even if the deformed set of generating functions is orthogonal the angular momentum projected set is not orthogonal. Fortunately, an orthogonal set of projected states is obtained if we chose an appropriate single particle state factor. 

Thus, it can be proved that  the following subset of projected states is orthogonal:
\bea
\phi_{IM;\sigma}\left(nl;d\right)&=&{\cal N}_{nl}^I(d)\left[P_{MI}^I|nlI\rangle
\Psi_c(d)\right]\chi_\sigma, \; \; {\rm{for}} \;I\neq 0,\; l=even,
\nonumber\\
\phi_{00;\sigma}(nl;d)&=&{\cal N}_{nl}^0(d)\left[P_{00}^0\left[|nl\rangle\hat{s}
\right]_{l+1,0} \Psi_c(d)\right]\chi_\sigma,~~{\rm{for}}\;I=0,\; l=odd,
\eea
where $\hat{s}$ denotes the spin operator and $\chi_{\sigma}$ is the bi-spinor component. The standard notation for the angular momentum projection operator was used:
\be
P_{MK}^I=\frac{2I+1}{8\pi^2}\int D^{I*}_{MK}(\Omega)\hat{R}(\Omega)d\Omega.
\ee
The norms ${\cal N}^{I}_{nl}(d)$ of these projected states are
\bea
\left[{\cal N}^{I}_{nl}\right]^{-2}&=&\sum_{J}\left(C^{l\,J\,I}_{I\,0\,I}\right)^{2}\left(N_{j}^{(c)}\right)^{-2}, \; \; {\rm{for}} \;I\neq 0,\; l=even,
\nonumber\\
\left[{\cal N}^{0}_{nl}\right]^{-2}&=&\frac{1}{4}\frac{1}{2l+3}\left(N_{l+1}^{(c)}\right)^{-2},~~{\rm{for}}\;I=0,\; l=odd,
\eea
 with $N^{(c)}_J$ denoting the norm of the $J$ component projected from the deformed state $|\Psi_c\rangle$, describing the core.
If one neglects the matrix elements with $\Delta l=\pm2$ and $\Delta n=2$,
the eigenvalues of $H$ within the projected spherical basis
can be fairly well approximated by the average values:
\bea
\nonumber
\epsilon_{nl}^I(d)&\equiv&<\phi_{IM;\sigma}(nl;d)|H|\phi_{IM;\sigma}(nl;d)>=
\hbar \omega_0 \left(N+\frac32\right)-D\left[l(l+1)-\frac{N(N+3)}{2}\right]\\
&+&\hbar \omega_0
\left(N+\frac32\right)\frac{1}{90}\left(\Omega_{\perp}^2-\Omega_z^2\right)^2
\left[1+\frac{1}{d^2} \langle\sum\limits_\mu b^{\dagger}_{2\mu} b_{2\mu}\rangle\right]\\
\nonumber
&-&\hbar\omega_0 \left(N+
\frac32\right)\frac13 \left( \Omega_{\perp}^2-\Omega_z^2\right) F_{Il}.
\eea

The expected value of the boson number operator as well as the factor $F_{Il}$ were calculated analytically in Ref.\cite{Rad}. The energies $\epsilon_{nl}^I(d)$ depend on $d$ in the same manner as the single particle energies in the Clemenger's model depend on the deformation $\delta$. This comparison has been performed in Ref.\cite{Rad}. The deformation parameter is fixed so that the cluster energy is minimum against any variation of $d$.

As shown in Ref. \cite{Rad3}, although the functions are characterizing a particle-core system, they can be used to describe a many fermion system. Indeed, in calculating the matrix elements of an one body operator, one integrates first on the core coordinates and then on single particle coordinates. Finally, the matrix element is written in a factorized form, one factor describing the matrix elements between spherical wave functions and the other one carrying the dependence on the deformation. Moreover, due to the specific properties of the coherent state the projected spherical states can be used for calculating the matrix elements of a two body operator. Indeed, in Ref.\cite{Rad4} we have proved that the matrix elements of a two body interaction are practically equal to the matrix elements between states projected  from the product of two particle states and  a common core's coherent state.

To conclude the single particle energies are approximated by those given in Eq. (2.13) while the corresponding wave functions by Eqs. (2.10), (2.12). Actually, these are input data for the treatment of a many body Hamiltonian associated to a system of interacting valence electrons
moving in the  mean field presented above.

\renewcommand{\theequation}{3.\arabic{equation}}
\section{RPA description of the collective dipole states}
\label{sec: level3}

We assume that the valence electrons moving in the mean field of the ionic core and interacting among themselves through a Coulomb force are described by the model Hamiltonian
\be
H=H_{mf}+V_{C},
\ee
which is a sum of the static self-consistent, single particle Hamiltonian $H_{mf}$
and of the two-body residual Coulomb interaction $V_{C}$. The residual two-body interaction is given, in the local density approximation, by
\be
V(\vec{r}_{1}-\vec{r}_{2})=\frac{e^{2}}{|\vec{r}_{1}-\vec{r}_{2}|}+\frac{dV_{XC}[\rho]}{d\rho}\delta{(\vec{r}_{1}-\vec{r}_{2})}.
\ee
The quantity $V_{XC}=d\varepsilon_{XC}[\rho]/d\rho$ is the exchange-correlation potential in the ground state. We use the exchange-correlation energy density $\varepsilon_{XC}$ of Gunnarsson and Lundqvist \cite{Gunn} as in Refs. \cite{Eckard,Brack}. Thus, the following expression for the exchange-correlation potential $V_{XC}$  given in atomic units, is obtained 
\be
V_{XC}(\vec{r})=-\frac{1.222}{r_{s}(\vec{r})}-0.0666\ln{\left(1+\frac{11.4}{r_{s}(\vec{r})}\right)}.
\ee
Here $r_{s}(\vec{r})=\left[3/4\pi\rho(\vec{r})\right]^{1/3}$ is the local value of the Wigner-Seitz radius.
The two body interaction is expanded in multipoles, the $\lambda$-pole term having the expression
 \cite{Eckard}:
\be
V(r_{1},r_{2};\lambda)=e^{2}\frac{r_{<}^{\lambda}}{r_{>}^{\lambda+1}}+\frac{dV_{XC}[\rho]}{d\rho}\frac{\delta{(r_{1}-r_{2})}}{r_{1}^{2}}\frac{2\lambda+1}{4\pi},
\ee
where $r_{<}=\min(r_{1},r_{2})$ and $r_{>}=\max(r_{1},r_{2})$.  

For $\lambda=1$, the two body interaction has the expression:
\bea
V(r_1,r_2;\lambda =1)&=&e^2\frac{r_1}{r_2^2}+F(\rho,r_s)\frac{\delta(r_1-r_2)}{r_1^2},\nonumber\\
F(\rho,r_s)&=&-\frac{1}{4\pi\rho r_s}\left[1.222+\frac{0.759r_s}{r_s+11.4}\right].
\eea

The one body term represented by the mean field $H_{mf}$ and the $\lambda=1$-pole term of the two body interaction are treated within the RPA formalism which defines an operator 
\be
C^{\dagger}_{1\mu}=\sum_{ph}\left[X_{ph}(c_{p}^{\dagger}c_{h})_{1\mu}-Y_{ph}(c_{h}^{\dagger}c_{p})_{1\mu}\right],
\label{phononop}
\ee
subject to the restrictions:

\bea
\left[H,C^{\dagger}_{1\mu^{\prime}}\right]&=&\hbar\omega C^{\dagger}_{1\mu^{\prime}},
\label{Harm}
\\
\left[C_{1\mu},C^{\dagger}_{1\mu^{\prime}}\right]&=&\delta_{\mu,\mu^{\prime}}.
\label{normalph}
\eea
Written in a matricial form, 
the RPA equations provided by (\ref{Harm}) look like: 
\be
\left(\begin{array}{cc}
A&B\\
-A^{*}&-B^{*}\end{array}\right)\left(\begin{array}{c}X^{n}\\
Y^{n}\end{array}\right)=\hbar\omega_{n}\left(\begin{array}{c}X^{n}\\
Y^{n}\end{array}\right).
\label{RPAeq}
\ee
The  sub-matrices $A$ and $B$ have the expressions: 
\bea
A(ph;p'h')&=&(\varepsilon_{p}-\varepsilon_{h})\delta_{p,p'}\delta_{h,h'}-B(ph;p'h'),\nonumber\\
B(ph;p'h')&=&2\frac{\hat{I}_{p}\hat{I}_{h}\hat{l}_{h}\hat{I}_{p'}\hat{I}_{h'}\hat{l}_{h'}}{3}C^{l_{h}\,1\,l_{p}}_{0\,0\,0}C^{l_{h'}\,1\,l_{p'}}_{0\,0\,0}f_{l_{p},l_{h}}^{I_{p},I_{h}}(d)f_{l_{p'},l_{h'}}^{I_{p'},I_{h'}}(d)R(ph';hp'),
\label{AsiB}
\eea
where
\be
R(ph';hp')=\int r_{1}^{2}dr_{1}r_{2}^{2}dr_{2}\mathcal{R}_{n_{p}l_{p}}(r_{1})\mathcal{R}_{n_{h'}l_{h'}}(r_{2})V(r_{1},r_{2};1)\mathcal{R}_{n_{h}l_{h}}(r_{1})\mathcal{R}_{n_{p'}l_{p'}}(r_{2}),
\ee
with $\mathcal{R}_{n_{i}l_{i}}(r)$ being the radial part of the single particle projected wave functions. Since the two-body interaction consists of two terms, the Coulomb and the exchange term, correspondingly the factor $R(ph';hp')$ split into two parts which are given in Appendix A.   The factor 2 of the sub-matrix $B$ accounts for the spin degeneracy. 

The factor $f(d)$ carrying the dependence on deformation parameter $d$, involved in Eq.(\ref{AsiB}),  have the expressions given in Appendix A.
Eq.(\ref{RPAeq}) determine the amplitudes $X$ and $Y$ up to a multiplicative constant which is fixed by the normalization condition (\ref{normalph}):
\be
\sum_{ph}\left[|X^{n}_{ph}|^2-|Y^{n}_{ph}|^2\right]=1.
\ee
The compatibility condition for the set of equations (\ref{RPAeq}) is an equation for $\omega$.      The number of solutions, for the RPA equations, is equal to the number of  dipole particle-hole ($ph$) configurations, hereafter denoted by $N_s$. To the solution $\hbar\omega_k$, the amplitudes $X^k_{ph}$ and $Y^k_{ph}$ correspond. These amplitudes characterize the phonon operator $C^{\dagger}_{1\mu}(k)$ which may excite the cluster ground state $|0\rangle$ to a one phonon state:

\be
C_{1\mu}(k)|0\rangle =0,\;\; |1_k\mu\rangle =C^{\dagger}_{1\mu}(k)|0\rangle .
\ee

There are several procedures to solve numerically the RPA equations. Here we have adopted the method proposed by Rowe in Ref.\cite{Rowe}.

Concerning the RPA description, we would like to comment on the following features:

a) Due to its  specific structure, the state with $I=0$ and $l=odd$  may be related, by the $ph$  dipole operator, to  states with either the spin up or the spin down. Consequently the RPA matrix has not a block structure, each block being characterized by an unique orientation of the electron spin. The coupling terms are however small and bring negligible contribution. For this reason we ignore from the beginning the terms which flip the spin of the $ph$ matrix elements.

b) In the definition of the phonon operator (\ref{phononop}) the summation involves both
states (2.10) with the factor state $\chi_{1/2}$ and with the bi-spinor $\chi_{-1/2}$. We, conventionally, call the resulting operator as {\it the extended phonon operator}.
In virtue of a) we may restrict the summation to one component and therefore work with the so called {\it reduced phonon operator}.
The normalization to unity of the two phonon operators suggest a simple relationship between their
defining amplitudes.

In our approach the reduced phonon operator has been used, otherwise the degeneracy of the states with spin up and with spin down has been carefully implemented whenever the  matrix elements between RPA states were calculated.

\renewcommand{\theequation}{4.\arabic{equation}}
\section{E1 transitions and photoabsorbtion cross section}
\label{sec: level4}

The reduced probability for the dipole transition $|0\rangle \to |1_{n}^{-}\rangle$ can be written as $^{[1]}$. \setcounter{footnote}{1}\footnotetext{ Throughout this paper the Rose's convention for the reduced matrix elements are used.}

\be
B(E1,0^{+}\rightarrow1_{n}^{-})=\left|\langle0||{\cal M}(E1)||1_{n}^{-}\rangle\right|^{2},
\ee
where
\be
\langle0||{\cal M}(E1)||1_{n}^{-}\rangle=\sum_{ph}\hat{I}_{p}\langle p||{\cal M}(E1)||h\rangle\left[X_{ph}^{n}+(-)^{I_{p}+I_{h}}Y_{ph}^{n}\right]
\ee
are the reduced matrix elements of the dipole operator ${\cal M}(E1)$, between the specified RPA state.
The one  phonon state $|1_n\rangle$ is characterized by the RPA amplitudes  $X_{ph}^{n}$ and $Y_{ph}^{n}$, obtained by solving Eq.(3.7). 

Instead of the usual transition dipole operator, we make use of a modified operator which is similar to the so-called Schiff moment from nuclear physics \cite{Zelev}.
\be
{\cal M}(E1)=\sqrt{\frac{4\pi}{3}}e\mathcal{Y}_{1\mu}(\Omega)\left(r-\frac{3}{5}\frac{r^{3}}{r_{s}^{2}}\right).
\ee
Here $r_{s}$ is the Wigner-Seitz radius and is equal to 3.93 (a.u.) for Na clusters.
The corrective component, involved in the dipole operator, relates particle and hole states characterized by $\Delta N =3$, which results in modifying the strength distribution among the RPA states. Such an effect is obtained in a natural manner, i.e. using the standard form for the dipole transition operator, if the mean field potential for the single particle motion involves higher powers of the radial coordinate.

We define the oscillator strength $f_{n}$ per atom as
\be
f_{n}=\frac{\hbar\omega_{n}B(E1,0^{+}\rightarrow1_{n}^{-})}{S(E1)},
\ee
where $\omega_{n}$ is the RPA excitation energy corresponding to the n-th order solution of RPA equation, and
\be
S(E1)=\sum_{n}\hbar\omega_{n}B(E1,0^{+}\rightarrow1_{n}^{-}),
\ee
such that $\sum_{n}f_{n}=1$.

To calculate the photoabsorbtion cross section per atom, $\sigma(\omega)$, one folds the oscillator strengths, which are just vertical straight lines,  with Lorentzian shapes normalized to unity as follows:
\be
\sigma(\omega)=C\sum_{n}f_{n}L(\omega;\omega_{n},\Gamma_{n}),
\label{sigmaofom}
\ee 
where $\hbar\Gamma_{n}$ denotes the full widths at half maximum of the Lorentzian profiles, and is provided by fixing the damping factor $\gamma=\Gamma/\omega_{r}$ with $\hbar\omega_{r}$ being the energy of the resonance peak. This damping factor varies, in our calculation, in the range of 0.06-0.135 \cite{Pacheco} which is appropriate for the room temperature. Indeed, the damping factor is considered to be caused by the coupling of the electronic dipole oscillation to the thermal fluctuations of the cluster surface \cite{Bertsch,Pacheco}. The thermal mechanism of broadening the plasmon line was first considered in Ref.\cite{Bertsch}. Therein the plasmon line is a Gaussian while in Ref.\cite{Pacheco} the plasmon profile is described by a Lorenzian function.
Note that the RPA calculations provides a line broadening due to the fragmentation of the collective strength onto near-lying excitations of a one-electron one-hole nature. This effect is temperature independent and is referred to as the Landau damping.

 The proportionality coefficient $C$,
in Eq. \ref{sigmaofom}, is given by 

\be
C=\frac{2\pi^{2}e^{2}\hbar}{m_{e}c}=1.0975\,(eV \AA^{2}).
\ee 
This value is obtained by normalizing the photoabsorbtion cross section such that the area per de-localized electron under the photoabsorbtion curve is constant \cite{Yann},
\be
\int_{0}^{\infty}\sigma({\omega})d(\hbar\omega)=\frac{2\pi^{2}e^{2}\hbar}{m_{e}c},
\ee
which is consistent with the value of the dipole sum rule.
Note that the results for the reduced E1 transition  probability as well as for the photoabsorbtion cross section depend on the single particle features specified by 
the mean field parameter $\hbar \omega_0$. The oscillator parameter of energy quanta 
$\hbar\omega_0$, depends on the cluster atoms number  as $E_{F}/{\cal N}^{1/3}$, where $E_{F}$ is the Fermi energy, which is about 3 eV for a spherical cluster. However, in general, $E_{F}$ has not a constant value for all clusters. Keeping this in mind, we use  a Fermi energy $E_{F}({\cal N})$ which  also depends on the number of cluster's atoms. The ${\cal N}$ dependence is extracted by interpolating the results provided by a least square fit for the experimental photoabsorbtion cross section spectrum. In our calculations one also needs to know the oscillator length $b$. This depends on the choice of oscillator energy quanta, and therefore exhibits the  ${\cal N}$ dependence given by $b=\left[\hbar^2/m_{e}E_{F}({\cal N})\right]^{1/2}{\cal N}^{1/6}$.

\renewcommand{\theequation}{5.\arabic{equation}}
\section{Electric polarizability}
\label{sec: level5}

In the classical picture, the static electric polarizability of a jellium metal sphere of radius $R$ has the expression:
\be
\alpha_{0}=R^{3},
\ee
where $R$, given in terms of the Wigner-Seitz radius $r_{s}$, is $R=r_{s}{\cal N}^{1/3}$.

Quantum mechanical effects determine corrections to the classical results for plasmon energy and polarizabilities. The electron density is not going sharply to zero at the cluster surface but reduces gradually at the surface and moreover extends significantly beyond to jellium edge. The spill-out electrons produce a screening effect against external fields which results in changing the classical result for polarizability to:
\be
\alpha=(R+\Delta)^{3}.
\ee 
The radius shift $\Delta$ can be expressed in terms of the fraction of the total number of electrons which are spilled-out the jellium sphere and the final result for the static polarizability reads:
\be
\alpha=R^{3}\left(1+\frac{\mathcal{N}_{sp}}{\mathcal{N}}\right),
\ee
where $\mathcal{N}_{sp}$ denotes the number of spilled-out electrons. Actually, this is a reasonable approximation of the result predicted by the sum rule $S_{-2}$\cite{Kresin,Giai,Brack1},
\be
\alpha=R^{3}\left(1-\frac{\mathcal{N}_{sp}}{\mathcal{N}}\right)^{-1}.
\ee
This expression is obtained by using the relationship between the plasmon redshift and the electric polarizability, provided by the moment $S_{-2}$. Also one assumes that the entire oscillator strength is concentrated near the surface plasmon, i.e. one ignores the presence of the volume plasmon, which allows us to use the result for the Thomas-Reiche-Kuhn sum rule $S_0$.

We propose a method for obtaining the number of spill-out electrons by means of the RPA formalism, using the RPA eigenstates for computing the average value of number of particles operator $\hat{N}=\sum_{i}c^{\dagger}_{i}c_{i}$, where the summation run over all particle and hole states. Thus the operator to be averaged is given by
\be
\hat{N}=\sum_{p}c^{\dagger}_{p}c_{p}+\sum_{h}c^{\dagger}_{h}c_{h}.
\ee
The averaging operation will be constrained by the condition that the radial integrals from all scalar products involved will have the limits $[R,\infty)$, instead of  $[0,\infty)$, where $R=r_{s}\mathcal{N}^{1/3}$ is the radius of the metal sphere which is supposed to contain all  electrons. In this manner we will get the average number of electrons which are beyond this sphere. The particle number operator can be written in a second quantization like form
\be
\hat{N}=\sum_{kk'}\langle C_{1\mu}(k)|\hat{N}|C_{1\mu}^{\dagger}(k')\rangle C_{1\mu}^{\dagger}(k)
C_{1\mu}(k'),
\ee

Averaging $\hat{N}$ with the RPA eigenstates we get:
\be
\langle\hat{N}\rangle=\sum_{k}\langle C_{k}|\hat{N}|C_{k}^{\dagger}\rangle=2\sum_{k}\sum_{ph}\left[\left(\nu(I_{p})X_{ph}^{k}f_{p}^{R}\right)^{2}+\left(\nu(I_{h})Y_{ph}^{k}f_{h}^{R}\right)^{2}\right].
\ee 
Here,  the overlap of two single particle projected functions corresponding to two particles or two holes states are expressed through the product of the statistical factor $\nu(I_{i})$ and the radial integral
\be
f_{i}^{R}=\int_{R}^{\infty}\left[\mathcal{R}_{n_{i}l_{i}}(r)\right]^{2}r^{2}dr,\,\,i=p,h.
\ee

\renewcommand{\theequation}{6.\arabic{equation}}
\section{Numerical results}
\label{sec: level6}

\begin{figure}[h!]
\begin{center}
\includegraphics[width=0.75\textwidth]{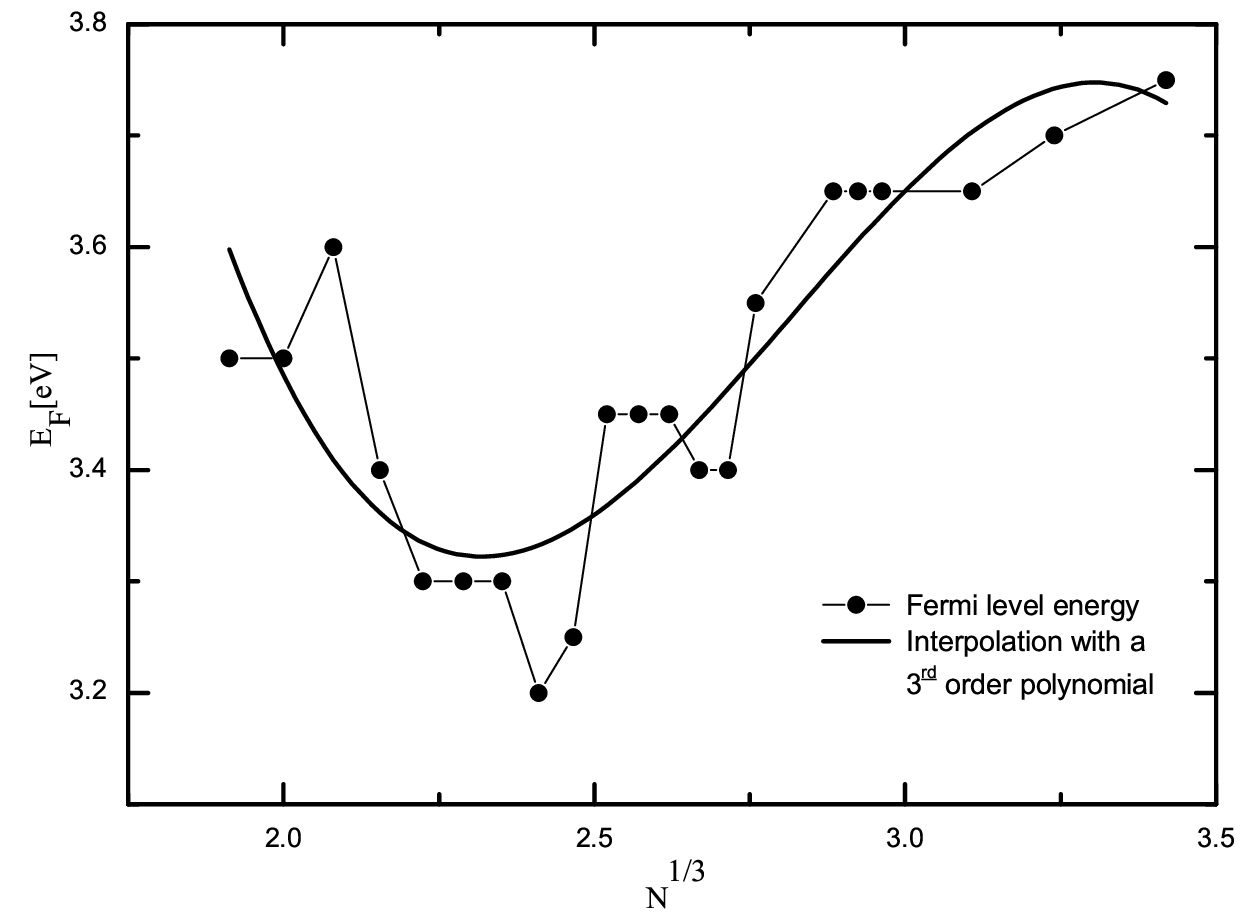}
\end{center}
\caption{The Fermi energies (black circles) which reproduce the experimental photoabsorbtion cross section spectra, are interpolated with a third order polynomial in $\mathcal{N}^{1/3}$: $E_{F}(\mathcal{N})=-0.886466{\cal N}+7.47916{\cal N}^{2/3}-20.3871{\cal N}^{1/3}+21.4339$.}
\label{Fig. 1}
\end{figure}

First, we identify the dependence of the Fermi energy on the number of atoms in cluster. 
The adopted procedure is as follows: For each cluster we determine the Fermi energy  which
corresponds to the best agreement of the calculated photoabsorbtion curve with the experimental points. Further, the obtained values are interpolated with a third order polynomial in $\mathcal{N}^{1/3}$ (see Fig.\ref{Fig. 1}). The polynomial obtained in this way determines a Fermi energy varying in the range of 3.3-3.75 eV. The RPA calculations make use of the Fermi energies lying on the interpolating curve. In this way  the  RPA results depend on the number of atoms, by means of the oscillator energy quanta $\hbar\omega_0=E_{f}(\mathcal{N})/\mathcal{N}^{1/3}$ and the oscillator length $b=\sqrt{\hbar/m_{e}\omega_0}$.

\begin{figure}[h!]
\begin{center}
\begin{tabular}{cc}
\epsfig{file=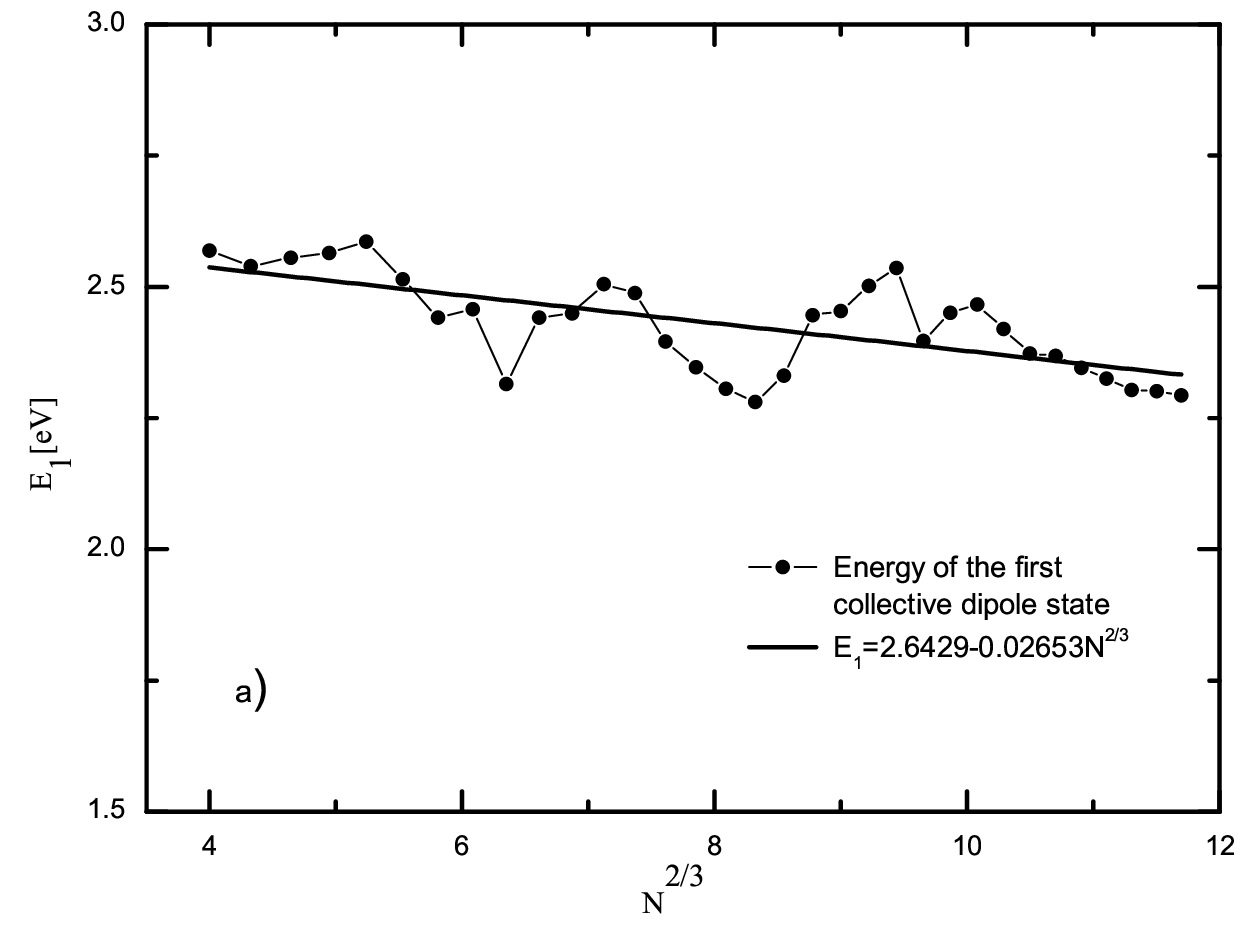,width=0.5\linewidth}
&\epsfig{file=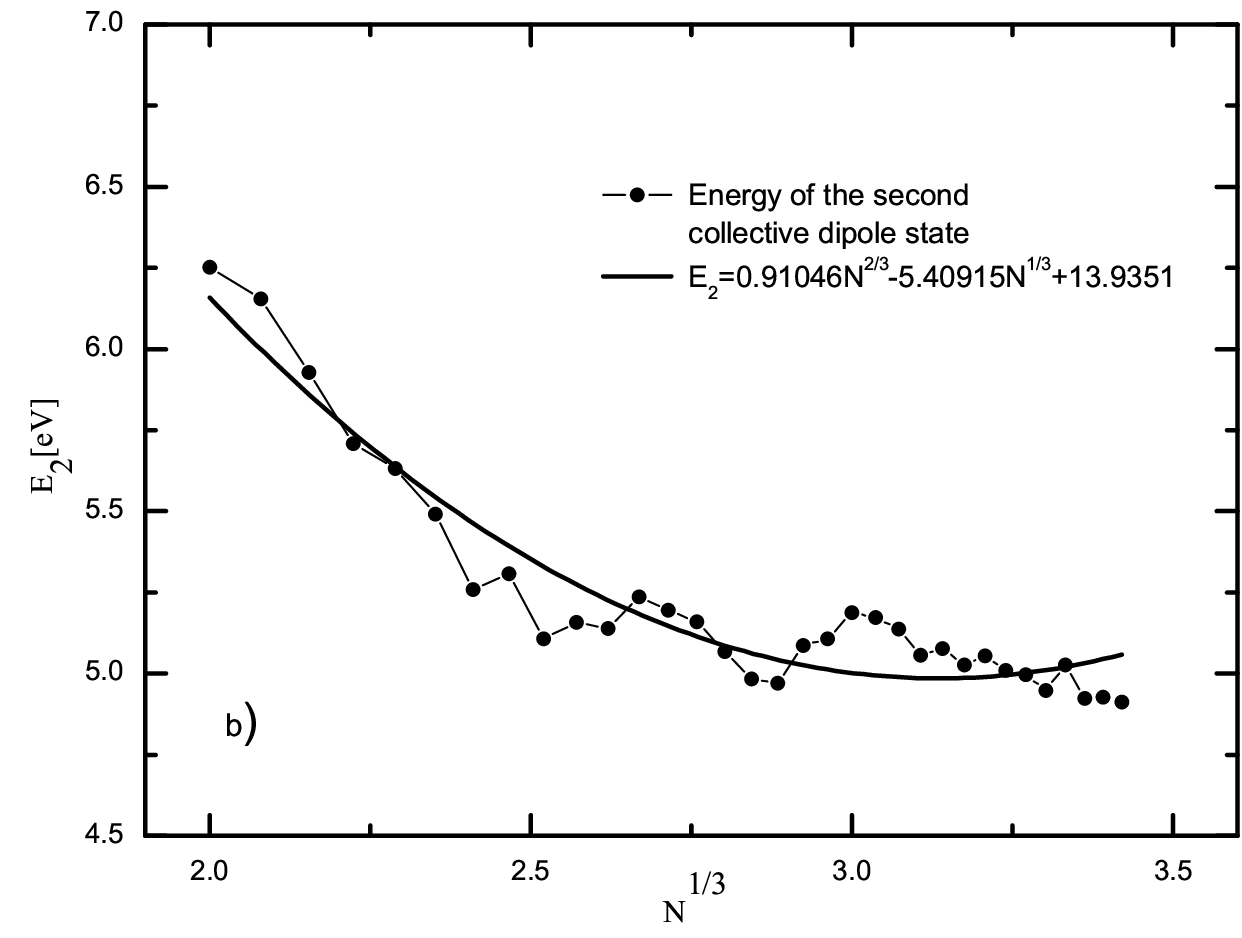,width=0.5\linewidth}
\end{tabular}
\end{center}
\caption{The RPA energy of the first collective dipole state decrease very slowly and almost linearly in $\mathcal{N}^{2/3}$ (a). For the second collective dipole state the energy is decreasing faster by means of a parabolic law in $\mathcal{N}^{1/3}$ (b). Theoretical results for all the clusters in the range of 8-40 atoms constituents, are presented.}
\label{Fig. 2}
\end{figure}

The RPA calculations are performed for the restricted subspace of the particle-hole excitations, including only the $\triangle N=1$ and $\triangle N=3$ transitions. We use the method described in Ref. \cite{Rowe} to solve the RPA equations with real sub-matrices $A(ph;p'h')$ and $B(ph;p'h')$. The single particle energies $\varepsilon_{i}$ involved in the expression of the sub-matrix $A(ph;p'h')$ are provided by the formula (2.13) in units of $\hbar\omega_0$. 
Once the RPA amplitudes and energies are  determined, the electric dipole transition probabilities and the oscillator strengths can be  computed by means of Eqs. (4.1) and (4.4). The biggest values for the transition probabilities define the  collective states. The corresponding
 RPA energies are located  mainly in two regions associated to the first collective dipole state, with the major contribution coming from the $\triangle N=1$ excitations, and the second collective dipole state, which is due to the $\triangle N=3$ transitions. The energy of the state characterized by a dominant transition probability varies from one cluster to another. Thus, the energies of the first and second collective states, depend on
$\mathcal{N}^{2/3}$  and $\mathcal{N}^{1/3}$ respectively, as shown in Fig.\ref{Fig. 2}.
Therefore, the first collective state seems to have a surface mode behavior, while the second collective excitation exhibits both a surface and a volume feature. In a short interval of $\mathcal{N}$ the volume character of the collective dipole mode of larger energy, prevails. 

In Table I, we collected the RPA amplitudes $X_{ph}$ and $Y_{ph}$ associated to the first collective dipole states in some  Na clusters. From there one sees that notable contributions are brought by several particle-hole configurations, which contrasts the situation of the non-collective states where a certain component  is by far dominant, while the others are negligible.

\begin{figure}[h!]
\begin{center}
\includegraphics[width=0.95\textwidth]{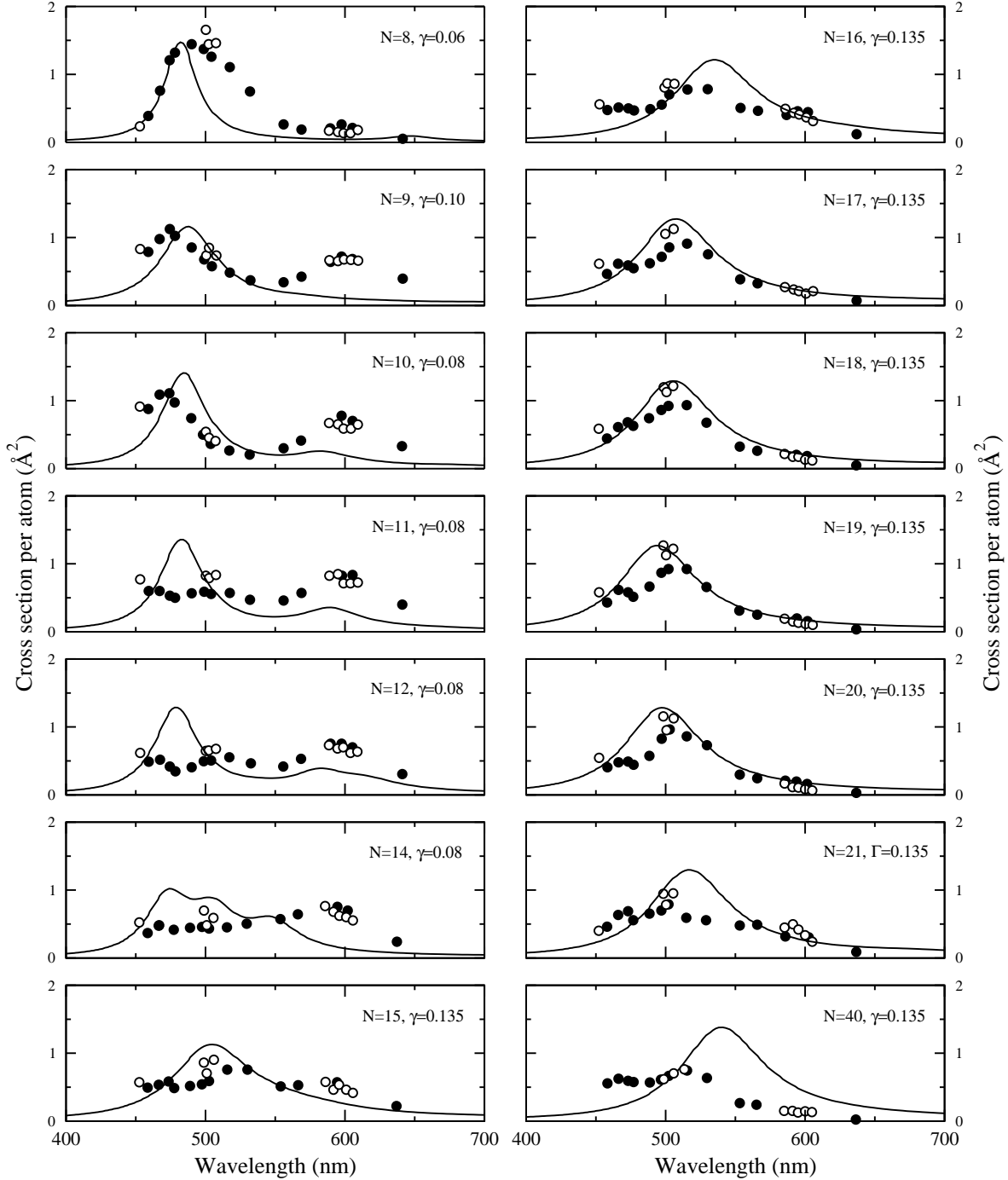}
\end{center}
\caption{Experimental (circles) and calculated (solid lines) photoabsorbtion cross sections vs wavelength for Na clusters of small and medium sizes. The experimental points are taken from ref. \cite{Selby1}, where the open circles corresponds to data taken with a flash-lamp laser and have about $20\%$ statistical errors. Solid circles have smaller statistical errors, about $5-10\%$, for which cw lasers were used. The calculated curves are resulting from the folding of the RPA oscillator strengths with Lorentzian shapes normalized to unity. The Lorentzian shapes are specified by the relative width parameter $\gamma=\Gamma/\omega_{r}$, which varies from one cluster to another in the range of 0.06 to 0.135, $\hbar\omega_{r}$ being the peak energy of a given Lorentzian.}
\label{Fig. 3}
\end{figure}

In order to compare the predictions of the present formalism with the experimental results for photoabsorbtion cross sections, we have also to  take into account the coupling of the electronic dipole oscillations with the thermal fluctuations of the cluster surface 
\cite{Bertsch,Pacheco,Voll}. In the generalized picture, the photoabsorbtion cross section per atom for a spherical cluster that is much smaller than the photon wavelength \cite{Mie}, have the plasmon energy dependence
\be
\sigma(\omega)=4\pi\frac{e^{2}}{m_{e}\hbar c}\frac{\omega^{2}\Gamma}{\left[\omega^{2}-\omega_{r}^{2}\right]^{2}+\omega^{2}\Gamma^{2}},
\ee 
where $\Gamma$ is an averaging parameter \cite{Pacheco}. Here {\it the surface plasmon pole approximation} has been adopted.  The resonance energy was denoted by $\hbar\omega_r$ while $\hbar\Gamma$ is just the 
resonance width. Keeping close to the general description of the photoabsorbtion cross section, the above formula can be simulated approximately by folding one oscillator strength line, predicted by the RPA calculation with a Lorentzian function displaying a damping factor $\gamma=\Gamma/\omega_{r}\approx0.1$, which is appropriate for room temperature. This procedure is extended to folding all the strengths provided by the RPA calculation, according to Eq.(4.6). Indeed, to each energy $\hbar\omega_k$ one associates a Lorenzian centered in  $\hbar\omega_k$ and having a  width given by the product of a constant damping factor $\gamma$ and the Lorenzian resonance energy. In Fig.\ref{Fig. 3} theoretical curves of photoabsorbtion cross section per atom are displayed versus the wavelength, for several clusters  for which experimental data are known in a large interval. Note that the RPA energies were given in terms of the corresponding wavelength $\lambda=2\pi \hbar c/\omega_k$. Apart from some discrepancies in the form of the curves, like plateaus or number of peaks, the present RPA calculations provide an overall good agreement with the experimental measurements in the visible range of energies. An interesting shape of the photoabsorbtion cross section is obtained for $Na_{14}$ cluster, which has a triaxial shape reflected in the fragmentation of the oscillator strength spectrum into three comparable peaks. These three distinct peaks are identified as plasmon frequencies associated to those three axis which define the triaxiality of the cluster. In a spherical cluster like $Na_{8}$ and $Na_{20}$ these three frequencies are degenerate which results in having only one peak in the photoabsorbtion cross section curves. Also, our calculation predicts the double peak structure of photoabsorbtion spectra for clusters $Na_{10}$, $Na_{11}$ and $Na_{12}$. Actually this structure suggests an axially symmetric shape. 

The photoabsorbtion spectrum for the clusters $Na_N$, with $N\le 40$, has been measured in Refs.
\cite{Voll,Selby1}. Therein, the measured data are interpreted within the ellipsoidal shell model (ESM) formalism. Although, at the first glance, the curves presented in the quoted references and here look similarly, there are some differences which are to be mentioned. For $Na_{20}$ the shape is quite well reproduced by our formalism while the predictions for the peak high and energy given in Ref.\cite{Selby1} are quite different from the corresponding experimental data. Concerning $Na_8$, both formalisms predict a resonance energy which is smaller than the experimental one. In our approach the ascending branch of the experimental curve is well reproduced, but the descending one exhibits a deviation due to the small value for the relative width parameter. For $Na_{12}$ the first two experimental peaks correspond to one large peak, in our calculations. For $Na_{14}$ we predict a triaxial shape reflected in the three peaks shown in Fig. 3, while in Ref.\cite{Selby1} only one peak is noticed. For $Na_N$ with $17\le N\le 21$, our calculations describe the experimental data better than ESM. As for $Na_{40}$ both formalisms predict a broad line with a peak which is higher than that shown by the data. The centroid energy calculated in the present paper is higher than the measured resonance energy, while in Ref.\cite{Selby1} the centroid energy is smaller.
A possible cause for the differences mentioned above is the use of different single particle bases in the two approaches.    
 
\begin{figure}[ht!]
\begin{center}
\includegraphics[width=0.8\textwidth]{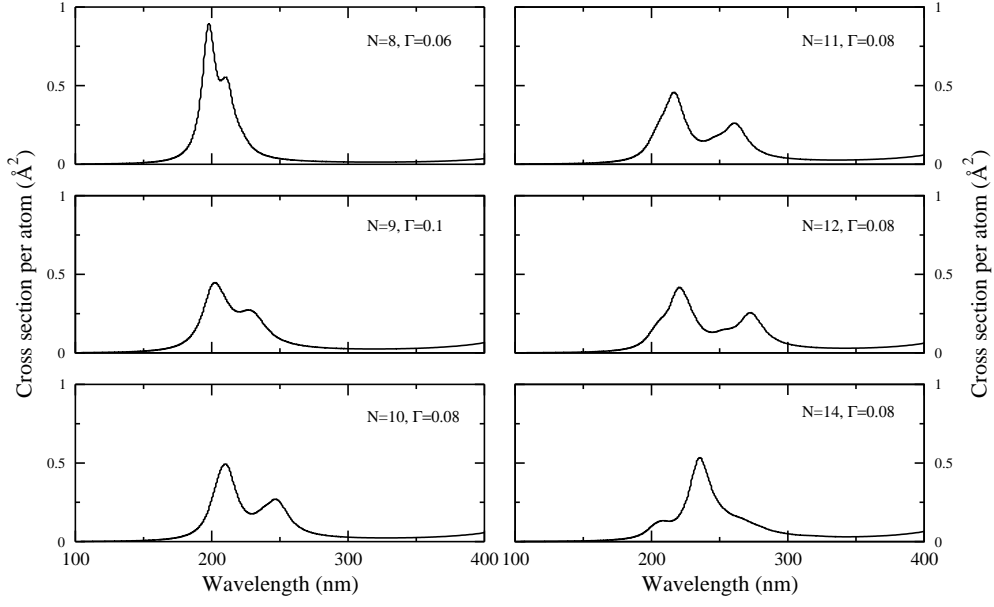}
\end{center}
\caption{Photoabsorbtion cross section vs wavelength for the second dipole collective state which can be associated to the volume plasma resonances. }
\label{Fig. 4}
\end{figure}

Taking a look at the photoabsorbtion cross section spectra in the ultraviolet region of wavelengths (Fig. ~\ref{Fig. 4}), where the second collective dipole state shows up, it is noticed that the corresponding photoabsorbtion peaks are also fragmented for some clusters. For some clusters there exists only one peak in this wavelength domain and this happens due to the near spherical form of these clusters. Indeed, the single peak shape is caused by the fact that for spherical clusters there is only one collective state, while for strongly deformed clusters this state is split into two parts, one consisting in two degenerate states of energy $\hbar\omega_{x}=\hbar\omega_{y}$, and another one of energy $\hbar\omega_{z}$. The ordering of the two energies depends on the cluster shape. For prolate clusters $\omega_{z}\leq\omega_{x}$, while for oblate clusters the ordering is changed. It is natural to suppose that the highest peak corresponds to the double degenerate energies $\hbar\omega_{x}=\hbar\omega_{y}$. Keeping this picture in mind and inspecting the crossections shown in Fig.\ref{Fig. 4}, one can conclude that for most  clusters with the second collective dipole states fragmented, the ordering of the larger and the smaller peaks suggest a   
prolate shape. By contrast, the cluster $Na_{14}$ has an oblate shape. This reasoning agrees perfectly well with the experimental data. Also, it is known that $Na_{8}$ is a spherical cluster, which is reflected in a single almost degenerate peak of the second collective dipole state.
The dependence of the plasmon profile on the cluster shape has been studied experimentally in Ref.
\cite{Borgg}. Also the aforementioned features are consistent with the semiclassical results
presented in Ref.\cite{Raduta}. Concluding, our analysis indicates that the cluster shape is reflected
in the profiles of both surface and volume plasmon. 

Comparing the surfaces covered by the curves in Fig.4 and by the corresponding ones from Fig.3 respectively, one 
may conclude that the volume plasmon takes only a small fraction of the dipole strength, which is in agreement with the result of Ref.\cite{Borgg} saying that the surface  plasmon exhausts about 70-100\% of the dipole sum rule. A detail analysis of the two types of plasmons, surface and volume, has been given in Ref. \cite{Kresin}. Thus, the two plasmons are the analogous to the Goldhaber-Teller and Steinwedel-Jensen modes of nuclear systems, respectively. The volume plasmon 
for large systems does not couple to the light and therefore cannot be populated. The reason is the fact that the light waves are transverse while plasma waves in infinite medium are longitudinal. In finite-sized particles, on the other hand, the volume mode can couple to light.
The  weight factors of the two modes contributions to the photoabsorbtion cross section have been measured for some sodium clusters in Refs.\cite{Selby1,Polak,Brec}. Thus, the weights for the surface plasmon in Na$_8$ and Na$_{20}$ are $0.7\pm0.05$ \cite{Selby1} and $0.7\pm 0.1$ \cite{Polak}, respectively. Another type of experiment \cite{Brec} predicts for  the weight of the surface plasmon in Na$_8$, the value
$0.64\pm 0.3$. The ratios of the aforementioned surfaces for the two small clusters are close
to the corresponding experimental results given above. Thus, we may say that our results agree with the calculations of Kresin \cite{Kres3} which predict that the photoabsorbtion strength in neutral spherical $Na$ clusters is shared between a surface- and a volume-plasma resonance.

\begin{figure}[ht!]
\begin{center}
\includegraphics[width=0.6\textwidth]{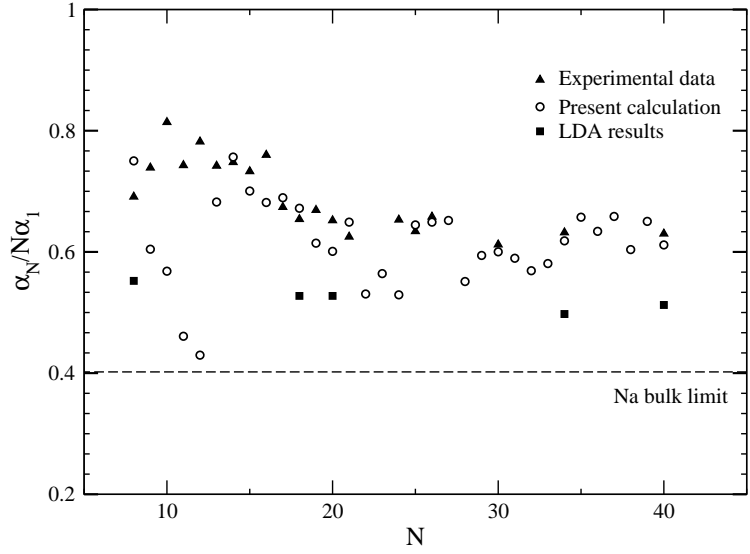}
\end{center}
\vspace*{-0.5cm}
\caption{The predicted static electric polarizabilities per atom for $Na_{\cal N}$ clusters normalized to the measured polarizability of Na atom (Ref. \cite{Molof}) (open circles), are compared with the corresponding experimental data (black triangles) from reference \cite{Voll} and with those given in Ref. \cite{Reimann} with a LDA approach. The bulk limit is also visualized.}
\label{Fig. 5}
\end{figure}

In Fig.\ref{Fig. 5} theoretical estimations for the static electric polarizability per atom, normalized to the value of polarizability of neutral Na atom are shown together with experimental values and LDA (Local Density Approximation) results, for a few clusters. The three data sets are  compared with each other as well as with the bulk limit result, which is associated to the classical polarizability of a metal-sphere. An excellent agreement is obtained for  most clusters' polarizabilities. However, there are
 few cases where big discrepancies are recorded, namely $Na_{9}$, $Na_{10}$, $Na_{11}$ and $Na_{12}$. The noteworthy fact is that  these clusters are most fragmented, some of them displaying a smooth two peak structure or even continuous plateaus of the photoabsorbtion cross section spectrum.   

It is fair to mention that the present work uses a temperature independent formalism. However, according to Ref. \cite{Blund} the temperature effect on the polarizability is relatively large. Indeed, the deviation of the measured polarizability at the room temperature from that corresponding to  T=0 is about 15\%. To this discrepancy one should add a correction due to the cluster shape thermal flutuation which amounts to $\pm 3\%$. If one adds the temperature effects
to the values plotted in Fig. 5, the agreement  between the calculated values and the corresponding data will be improved for some clusters and moderately altered for the remaining ones.

\vspace*{0.5cm}

\clearpage   

\begin{table}[ht!]
\caption{The dominant RPA  $X_{ph}$-amplitudes of the first collective dipole states achieved by  $\triangle N=1$ transitions for some of the Na clusters. The corresponding dipole $ph$ configuration as well the RPA energies are also mentioned. Note that the heavier are the clusters the more collective is the depicted RPA mode. Also due to the repulsive character of the two body interaction the order of the RPA root is increasing with ${\cal N}$ despite the fact that the energy is slightly decreasing.}
{\scriptsize
\begin{tabular}{|c|c|r@{.}l|r@{.}l|c|c|}
\hline
Cluster\raisebox{4.2ex}{~} & ~~~~~~~~~~~~~$[NlI]_{h}\,\longrightarrow\,[NlI]_{p}$~~~~~~~~~~~~~ & \multicolumn{2}{c|}{$X^k_{ph}$} & \multicolumn{2}{c|}{$Y^k_{ph}$} & $\begin{array}{c}\textrm{solution's}\\\textrm{order }k\end{array}$ &~~~$\hbar\omega_{k}$[eV]~~~\\
\hline
            & $[1\,1\,1]_{h}\,\longrightarrow\,[2\,2\,2]_{p}$ &~~~~~~~~~0&7154~~~~~~~~&~~~~~~~~~0&1525~~~~~~~~&    &\raisebox{3.5ex}{~}\\
 $Na_{8}$   & $[1\,1\,1]_{h}\,\longrightarrow\,[2\,0\,0]_{p}$ &     0&6297    &     0&0636    & 5  &       2.569       \\
            & $[1\,1\,0]_{h}\,\longrightarrow\,[2\,2\,1]_{p}$ &  $-$0&3416    &  $-$0&0723    &    &                   \\[1ex]
\hline
            & $[2\,2\,0]_{h}\,\longrightarrow\,[3\,1\,1]_{p}$ &     0&8795    &     0&0103    &    &\raisebox{3.5ex}{~}\\
            & $[2\,2\,1]_{h}\,\longrightarrow\,[3\,1\,1]_{p}$ &  $-$0&2628    &  $-$0&0131    &    &                   \\
            & $[1\,1\,1]_{h}\,\longrightarrow\,[2\,2\,2]_{p}$ &     0&2339    &     0&0611    &    &                   \\
 $Na_{14}$  & $[2\,2\,1]_{h}\,\longrightarrow\,[3\,3\,2]_{p}$ &     0&2044    &     0&0568    & 9  &       2.619       \\
            & $[2\,2\,0]_{h}\,\longrightarrow\,[3\,3\,1]_{p}$ &     0&1837    &     0&0600    &    &                   \\
            & $[1\,1\,1]_{h}\,\longrightarrow\,[2\,0\,0]_{p}$ &     0&1396    &     0&0384    &    &                   \\
            & $[2\,2\,1]_{h}\,\longrightarrow\,[3\,3\,1]_{p}$ &     0&1309    &     0&0514    &    &                   \\[1ex]
\hline
            & $[2\,2\,0]_{h}\,\longrightarrow\,[3\,3\,1]_{p}$ &     0&4709    &     0&1611    &    &\raisebox{3.5ex}{~}\\
            & $[2\,2\,2]_{h}\,\longrightarrow\,[3\,3\,3]_{p}$ &     0&4354    &     0&1492    &    &                   \\
 $Na_{18}$  & $[1\,1\,1]_{h}\,\longrightarrow\,[2\,0\,0]_{p}$ &     0&4213    &     0&0928    & 12 &       2.449       \\
            & $[2\,2\,1]_{h}\,\longrightarrow\,[3\,3\,2]_{p}$ &     0&4211    &     0&1441    &    &                   \\
            & $[2\,2\,1]_{h}\,\longrightarrow\,[3\,3\,1]_{p}$ &     0&3847    &     0&1315    &    &                   \\
            & $[2\,2\,2]_{h}\,\longrightarrow\,[3\,1\,1]_{p}$ &     0&3635    &     0&0635    &    &                   \\[1ex]
\hline
            & $[2\,0\,0]_{h}\,\longrightarrow\,[3\,1\,1]_{p}$ &  $-$0&5047    &  $-$0&1444    &    &\raisebox{3.5ex}{~}\\
            & $[2\,2\,0]_{h}\,\longrightarrow\,[3\,3\,1]_{p}$ &  $-$0&4536    &  $-$0&1621    &    &                   \\
            & $[2\,2\,2]_{h}\,\longrightarrow\,[3\,3\,3]_{p}$ &  $-$0&4194    &  $-$0&1501    &    &                   \\
 $Na_{20}$  & $[2\,2\,1]_{h}\,\longrightarrow\,[3\,3\,2]_{p}$ &  $-$0&4056    &  $-$0&1450    & 12 &       2.488       \\
            & $[2\,2\,1]_{h}\,\longrightarrow\,[3\,3\,1]_{p}$ &  $-$0&3705    &  $-$0&1323    &    &                   \\
            & $[2\,2\,2]_{h}\,\longrightarrow\,[3\,1\,1]_{p}$ &  $-$0&3432    &  $-$0&0657    &    &                   \\
            & $[2\,2\,1]_{h}\,\longrightarrow\,[3\,1\,0]_{p}$ &  $-$0&1623    &  $-$0&0312    &    &                   \\[1ex]
\hline
            & $[3\,3\,1]_{h}\,\longrightarrow\,[4\,4\,2]_{p}$ &  $-$0&4117    &  $-$0&1681    &    &\raisebox{3.5ex}{~}\\
            & $[3\,1\,1]_{h}\,\longrightarrow\,[4\,2\,2]_{p}$ &  $-$0&4075    &  $-$0&1369    &    &                   \\
            & $[3\,1\,1]_{h}\,\longrightarrow\,[4\,0\,0]_{p}$ &  $-$0&4040    &  $-$0&0993    &    &                   \\
            & $[3\,3\,2]_{h}\,\longrightarrow\,[4\,4\,3]_{p}$ &  $-$0&3182    &  $-$0&1298    &    &                   \\
 $Na_{40}$  & $[3\,3\,1]_{h}\,\longrightarrow\,[4\,2\,0]_{p}$ &  $-$0&2902    &  $-$0&0547    & 22 &       2.293       \\
            & $[3\,3\,2]_{h}\,\longrightarrow\,[4\,2\,1]_{p}$ &  $-$0&2593    &  $-$0&0489    &    &                   \\
            & $[3\,3\,1]_{h}\,\longrightarrow\,[4\,2\,1]_{p}$ &     0&2391    &     0&0446    &    &                   \\
            & $[3\,3\,0]_{h}\,\longrightarrow\,[4\,4\,1]_{p}$ &     0&2326    &     0&0942    &    &                   \\
            & $[3\,1\,0]_{h}\,\longrightarrow\,[4\,2\,1]_{p}$ &     0&1947    &     0&0648    &    &                   \\
            & $[3\,3\,2]_{h}\,\longrightarrow\,[4\,4\,2]_{p}$ &  $-$0&1673    &  $-$0&0689    &    &                   \\[1ex]
\hline
\end{tabular}}
\label{table 1}
\end{table}
\clearpage

\renewcommand{\theequation}{7.\arabic{equation}}
\section{Conclusions}
\label{sec: level7}
The main results described in the previous Sections can be summarized as follows.
The atomic clusters were replaced by a set of valence electrons moving in a mean field and 
interacting among themselves through a Coulomb and an exchange interaction. Although not presented analytically, the mean field is defined by a set of orthogonal projected spherical single particle states and a set of corresponding energies. The two-body interaction is expanded in multipoles, the expansion being truncated at $\lambda=1$. The mean field and the two-body  dipole interaction are treated within the RPA approach which defines a set of particle-hole like phonon states.
The E1 transition from the ground state to excited one phonon states were calculated. The collective E1 transitions to the low energy states, around 2.5 eV, are determined by a coherent contribution of the $\Delta N = 1$, $ph$ configuration. There are also states, of energies about 5-6 eV, which are collectively populated with $\Delta N=3$ $ph$ transitions. As shown in Fig. 2, the first collective mode has a surface character while the second one seems to be mainly of a volume type.

The photoabsorbtion cross section was obtained by folding the E1 strengths carried by the RPA one phonon states, with Lorentzian with a damping factor varying from one cluster to another, in the range of 0.06-0.135. The result shown in Eq.(4.6) is plotted in Fig.3 as a function of the wavelength associated to the RPA states, and compared with the experimental data.
The comparison reveals a reasonable good agreement. For the light clusters, N=9,10,11,12, the figures exhibit two peaks, which suggest the existence of two modes corresponding to oscillations along and perpendicular  to the symmetry axis, respectively. For N=14, the results of our calculation indicate a triaxial shape.
Similar fragmentation is also obtained  for the volume plasmon resonance, shown in Fig.4.
According to the relative magnitude of the two peaks, we concluded that the clusters with N=9,10,11,12 has a deformed prolate shape while $Na_{14}$ seems to be of an oblate shape. Perhaps a smaller damping factor would provide evidence for a triaxial shape also for the volume mode in
$Na_{14}$.

{\it It is noteworthy the fact that the fragmentation effects seen in both low (surface like) and high energy (volume like) modes cannot be described with a spherical single particle basis.}

Based on calculations for the number of spilled-out electrons, we  calculated the electric polarizability which is compared with the experimental data, the bulk limit result as well as with the LDA (local density approximation) predictions. Except for the clusters with N=9,10,11,12,24, the calculated polarizabilities agree quite well with the corresponding  experimental data. It is interesting to remark that concerning the clusters for which large discrepancies were recorded, some of predictions
(N=9,10,24) lie close to the LDA results while others (N=11,12) agree with the bulk limit of $Na$ clusters. Our formalism is a temperature independent approach. Adding the temperature effect the picture shown in Fig. 5 would be modified, at the room temperature, by about 15\% \cite{Blund}. However, the corrected polarizabilities are still in a reasonable good agreement with the experimental data.

The final conclusion is that the projected spherical single particle basis seems to be an useful tool for describing the many-body features of deformed atomic clusters.

\begin{acknowledgments}
This work was also supported  by the Romanian Ministry for Education and Research under the contract PNII, No. ID-33/2007.
\end{acknowledgments}

\renewcommand{\theequation}{A.\arabic{equation}}
\setcounter{equation}{0}

\section{Appendix A}
The two terms of the two body potential $V(r_1,r_2;1)$ determine two following two body matrix elements:
\bea
R(ph',hp')&=&R_1(ph',hp')+R_2(ph',hp'),\nonumber\\
R_1(ph',hp')&=&\int_{0}^{\infty}\left[\int_{0}^{r_2}r_1^3R_{n_pl_p}(r_1)R_{n_hl_h}(r_1)dr_1\right]
R_{n^{\prime}_pl^{\prime}_p}(r_2)R_{n^{\prime}_hl^{\prime}_h}(r_2)dr_2,\nonumber\\
R_2(ph',hp')&=&F(\rho,r_s)\int_{0}^{\infty}R_{n_pl_p}(r_1)R_{n_hl_h}(r_1)
R_{n^{\prime}_pl^{\prime}_p}(r_1)R_{n^{\prime}_hl^{\prime}_h}(r_1)r_1^2dr_1.
\eea

The factors $f$ involved in the RPA matrices have the expressions:
\bea
f_{l_{p},l_{h}}^{I_{p},I_{h}}(d)&=&-\frac{1}{2}\nu(I_{p})\nu(I_{h})\mathcal{N}_{n_{p}l_{p}}^{I_{p}}(d)\mathcal{N}_{n_{h}l_{h}}^{I_{h}}(d)\left[N_{l_{p}+1}^{(c)}\right]^{-2}\frac{1}{\sqrt{2l_{p}+3}}
\nonumber\\
&\times &C^{l_{h}\,1\,\,l_{p}}_{0\,\,\,\,0\,\,0}C^{l_{h}\,l_{p}+1\,I_{h}}_{I_{h}\,\,0\;\;\;\;\;\;I_{h}};W(1\;l_{p}\;I_{h}\;l_{h};\,l_{p}+1\;1),
\eea
for $(I_{p},l_{p})=(0,odd)$ and $(I_{h},l_{h})\ne(0,odd)$,
\bea
f_{l_{p},l_{h}}^{I_{p},I_{h}}(d)&=&\frac{1}{2}\nu(I_{p})\nu(I_{h})\mathcal{N}_{n_{p}l_{p}}^{I_{p}}(d)\mathcal{N}_{n_{h}l_{h}}^{I_{h}}(d)\left[N_{l_{h}+1}^{(c)}\right]^{-2}\frac{1}{\sqrt{2l_{h}+3}}
\nonumber\\
&\times &C^{l_{h}\,1\,\,l_{p}}_{0\,\,\,\,0\,\,0}C^{l_{p}\,l_{h}+1\,I_{p}}_{I_{p}\,\,0\;\;\;\;\;\;I_{p}}\;\;W(1\;l_{h}\;I_{p}\;l_{p};\,l_{h}+1\;1),
\eea
for $(I_{h},l_{h})=(0,odd)$ and $(I_{p},l_{p})\ne(0,odd)$,
\bea
f_{l_{p},l_{h}}^{I_{p},I_{h}}(d)&=&\nu(I_{p})\nu(I_{h})\mathcal{N}_{n_{p}l_{p}}^{I_{p}}(d)\mathcal{N}_{n_{h}l_{h}}^{I_{h}}(d)\nonumber\\
&\times& \sum_{J}C^{l_{p}\,J\,I_{p}}_{I_{p}\,0\,I_{p}}C^{l_{h}\,J\,I_{h}}_{I_{h}\,0\,I_{h}}\left[N_{J}^{(c)}\right]^{-2}W(I_{h}\,J\,1\,l_{p};\,l_{h}\,I_{p}),
\eea
for $(I_{p},l_{p})\neq(0,odd)$ and $(I_{h},l_{h})\neq(0,odd)$, where $\nu(I_{i})=\frac{2-\delta_{I_{i},0}}{2I_{i}+1}$ is the statistical factor connected to the fact that the energy level with a given $I$ have the degeneracy $2I+1$ and it contains also the  $I,-I$ degeneracy which add a factor $2$ if $I\ne 0$. The Clebsch-Gordon coefficients appearing in the above expressions take proper care of the angular momentum coupling, as well as of the parity-conservation conditions.


\begin{thebibliography}{99}

\bibitem{Knight}W. D. Knight, K. Clemenger, W. A. de Heer, W. A. Saunders,
M. Y. Chou and M. L. Cohen, Phys. Rev. Lett., {\bf 52}, 2141 (1984).

\bibitem{Gal} G. Galli and Parrinello, in Comp. Sim. in
Material Science, NATO ASI E: Applied Sciences, Vol. 205, eds.
 M. Meyer and V. Pontikis (Kluwer Academic, Dordrecht)p. 283.

\bibitem{Heer} W. A. de Heer, Rev. Mod. Phys. {\bf 65}, 611 (1993).

\bibitem{Brack1}M. Brack, Rev. Mod. Phys. {\bf 65}, 677 (1993).

\bibitem{Kresin} Vitaly V. Kresin, Physics Reports {\bf 220}, 1 (1992).

\bibitem{Cini} M. Cini, J. Catal., {\bf 37}, 187 (1975).

\bibitem{Bonac} V. Bonacic-Koutecky, P. Fantuccci and J. Koutecky,
Phys. Rev. {\bf B37}, 4369 (1988).

\bibitem{Kohn} W. Kohn and L. J. Sham, Phys. Rev. {\bf A140}, 1133 (1965).

\bibitem{Mart}J. L. Martins, J. Buttet and R. Cae, Phys. Rev. {\bf B31},
1884 (1985).

\bibitem{Beck} D. E. Beck, Solid State Communic. {\bf 49}, 381 (1984).

\bibitem{Nish} H. Nishioka, K. Hansen and B. Mottelson, Phys. Rev. {\bf
B42},9377 (1990).

\bibitem{Chou} M. Y. Chou and M. L. Cohen, Solid State Commun. {\bf 52},
645 (1984).

\bibitem{Voll} K. Selby, M. Vollmer, J. Masui, V. Kresin, W. A. de Heer and W. D. Knight, Phys. Rev. {\bf B 40} (1989) 5417.

\bibitem{Selby1}K. Selby, V. Kresin, J. Masui, M. Vollmer, W. A. de Heer, A. Scheidemann, W. D. Knight, Phys. Rev. B {\bf 43}, 4565 (1991).

\bibitem{Brec} C. Brechignac, Ph. Cahuzac, F. Carlier, M. de Frutos and J. Leygnier, Chem. Phys. Lett. {\bf 189}, 28 (1992).

\bibitem{Fall} H. Fallgren, K M. Brown and T. P. Martin, Z. Phys. D {\bf 19}, 81 (1991).

\bibitem{Yan}C. Yannouleas and R. A. Broglia, Phys. Rev. {\bf A44},
5793 (1991).

\bibitem{Yan1}C. Yannouleas and Uzi Landman, Phys. Rev. {\bf B51}, 1902 (1995).

\bibitem{Kos} M. Koskinen, P. O. Lipas and M. Manninen, 
Z. Phys. {\bf D35}, 285 (1995).

\bibitem{Clem} K. Clemenger, Phys. Rev. {\bf B32}, 1359 (1985).

\bibitem{Nils}S. G. Nilsson, K. Dan. Vidensk. Selsk. Mat. Fys. Medd.{\bf
29} No 16.
\bibitem{Rad}A. A. Raduta, Ad. R. Raduta, Al. H. Raduta, Phys. Rev. B {\bf 59}, 8209 (1999).

\bibitem{Pedersen} J. Pedersen, S. Bjornholm, J. Borgreen, K. Hansen, T. P. Martin
and H. D. Rasmussen, Nature, {\bf 353}, 733 (1991).

\bibitem{Martin} T. P. Martin, S. Bjornholm, J. Borgreen, C. Brechignac, Ph. Cahuzac, K. Hansen and J. Pedersen, Chem. Phys. Lett. {\bf 186}, 53 (1991).

\bibitem{Kosk1} M. Koskinen, P. O. Lipas and M. Manninen, Nucl. Phys.
{\bf A591}, 421 (1995).

\bibitem{Rigo}A. Rigo, M. Casas, F. Garcias, E. Moya de Guerra and P.
Sarriguren, Phys. Rev. {\bf B 57} (1998) 11943.  

\bibitem{Raduta}A. A. Raduta, E. Garrido, E. Moya de Guerra, Eur. Phys. J. D {\bf 15}, 65 (2001).

\bibitem{Rad3} A. A. Raduta, D. S. Delion and N. Lo Iudice, Nucl.Phys. {\bf A 551} (1993) 73.

\bibitem{Rad4} A. A. Raduta, A. Escuderos and E. Moya de Guerra, Phys. Rev. C 65 (2002)0243121.

\bibitem{Gunn}O. Gunnarsson, B. I. Lundqvist, Phys. Rev. B {\bf 13}, 4274 (1976).

\bibitem{Eckard}W. Eckard, Phys. Rev. B {\bf 29}, 1558 (1984).

\bibitem{Brack}M. Brack, Phys. Rev. B {\bf 39}, 3533 (1989).

\bibitem{Rowe}N. Ullah, D. J. Rowe, Nucl. Phys. A {\bf 163}, 257 (1970). 


\bibitem{Molof}R. W. Molof, H. L. Schwartz, T. M. Miller, B. Bederson, Phys. Rev. A {\bf 10}, 1131 (1974).

\bibitem{Reimann}S. M. Reimann, M. Koskinen, H. Hakkinen, P. E. Lindelof, M. Manninen, Phys. Rev. B {\bf 56}, 12147 (1997).

\bibitem{Zelev}V. Zelevinsky, A. Volya, N. Auerbach, Phys. Rev. C {\bf 78}, 14310 (2008).

\bibitem{Bertsch} G. F. Bertsch and D. Tomanek, Phys. Rev. B {\bf 40}, 2749 (1989).

\bibitem{Pacheco}J. M. Pacheco, R. A. Broglia, Phys. Rev. Lett. {\bf 62}, 1400 (1989).

\bibitem{Mie}G. Mie, Ann. Phys. (Leipzig) {\bf 25} (1908) 377.

\bibitem{Giai}N. Van Giai, Progress of Theoretical Physics, Supplement
No.124 (1996) 1.

\bibitem{Yann}C. Yannouleas, R. A. Broglia, Phys. Rev. A {\bf 44}, 5793 (1991).

\bibitem{Blund} S. A. Blundel, C. Guet and Rajendra R. Zope, Phys. Rev. Lett.{\bf 84}, 4826 (2000)

\bibitem{Borgg} J. Borggreen, P. Chowdhury, N. Kebaili, L. Lundsberg-Nielsen, K. \..{u}tzenkirchen, M. B. Nielsen, J. Pedersen and H. D. Rasmussen, Phys. Rev. B {\bf 48} 17507 (1993).

\bibitem{Polak} S. Pollack, C. R. C. Wang and M.M. Kappes, J. Chem. Phys. {\bf 94}, 2496 (1991).

\bibitem{Kres3}V. Kresin, Phys. Rev. {\bf B 42}, 3247 (1990).
\end{thebibliography}
\end{document}